\documentclass[a4paper,11pt]{article}
\usepackage{pos}

\newcommand{\be}{\begin{eqnarray}}
\newcommand{\ee}{\end{eqnarray}}

\title{A Space Mission to Earth's Nearest Black Hole:\\Reality or Science Fiction?}
\ShortTitle{A Space Mission to Earth’s Nearest Black Hole}

\author*[a,b]{Cosimo~Bambi}

\affiliation[a]{Center for Astronomy and Astrophysics, Department of Physics, Fudan University,\\
2005 Songhu Road, Shanghai 200438, China}

\affiliation[b]{School of Natural Sciences and Humanities, New Uzbekistan University,\\
Movarounnahr Street 1, Tashkent 100000, Uzbekistan}

\emailAdd{bambi@fudan.edu.cn}

\abstract{Black holes are the sources of the strongest gravitational fields in the present-day Universe, offering unparalleled opportunities to test Einstein's theory of General Relativity in the strong-field regime. In this talk, I will examine the prospect of sending small spacecraft on an interstellar mission to the nearest black hole. While highly speculative and fraught with technical challenges, such an endeavor is not entirely beyond the realm of possibility. Although the necessary technology does not yet exist, it may become available within the next 20 to 30 years. The mission itself could span 80 to 100 years, yet the scientific return would be profound and potentially unattainable through other means.}

\FullConference{Mondello Workshop 2026 Frontier Research in Astrophysics – V (FRAPWS2026)\\
15-20 June 2026\\
Palermo, Italy\\}

\begin{document}

\maketitle

%%%%%%%%%%%%%%%%%%%%%%%%%%%%%%%%%%%%%%%%%%%%%%%%%%

\section{Introduction}\label{s-intro}

The past $\sim$10 years have substantially transformed our understanding of black holes. Until 2015, observational evidence for astrophysical black holes remained indirect, and no tests confirmed that these objects behaved according to the predictions of General Relativity. We knew of stellar-mass black holes in X-ray binary systems and supermassive black holes in galactic nuclei. Stellar-mass black holes were recognized as compact objects exceeding the theoretical maximum mass for neutron stars~\cite{Rhoades:1974fn,Lattimer:2012nd}. Supermassive black holes, meanwhile, were deemed too massive, too compact, and too old to be clusters of neutron stars~\cite{Maoz:1997yd}. However, there was no evidence that the spacetime around these objects matched the Kerr solution predicted by General Relativity~\cite{Bambi:2015kza,Bambi:2017khi}.

Today, the situation is markedly different. Within the precision of current measurements, we can confirm that astrophysical black holes adhere to the predictions of General Relativity, drawing on gravitational-wave data (see, e.g., Refs.~\cite{LIGOScientific:2016lio,Yunes:2016jcc,LIGOScientific:2019fpa,Das:2026zyt}), X-ray observations of black hole binaries (see, e.g., Refs.~\cite{Cao:2017kdq,Tripathi:2018lhx,Tripathi:2020dni,Tripathi:2020yts,Bambi:2022dtw}), and the images of the supermassive black holes M87$^*$ and SgrA$^*$ (see, e.g., Refs.~\cite{EventHorizonTelescope:2020qrl,EventHorizonTelescope:2022xqj,Vagnozzi:2022moj}).

Future observational facilities will undoubtedly improve these tests and yield tighter constraints on possible deviations from General Relativity. Yet it is unlikely that we will ever achieve the level of precision and accuracy in testing black hole physics that we currently enjoy in, say, atomic or particle physics. The astrophysical environments hosting these objects are exceedingly complex and beyond our control, and we are unlikely to develop theoretical models sophisticated enough to analyze observations at the precision required for probing fundamental physics in a truly high-precision manner.

This raises a natural question: could we consider an interstellar mission to the nearest black hole, designed to perform extremely precise and accurate in-situ tests~\cite{Bambi:2025kcr,Bambi:2025hjn}? Although such a mission may sound like science fiction today, in this talk I will argue that it is not necessarily impossible, and that the requisite technology may become available within a few decades.

%%%%%%%%%%%%%%%%%%%%%%%%%%%%%%%%%%%%%%%%%%%%%%%%%%

\section{Population of Nearby Black Holes}\label{s-pop}

Today, fewer than 100~stellar-mass black holes are known in the Galaxy, and most of these reside in X-ray binaries; see, for instance, Ref.~\cite{Bambi:2025rod} for a review. However, black hole X-ray binaries are expected to be extremely rare systems. Most stellar-mass black holes in the Galaxy are thought to be isolated, without any companion.

Current astrophysical models suggest that the Galaxy hosts $10^8$-$10^9$~stellar-mass black holes formed from the gravitational collapse of stars with masses $M_{\rm star} \gtrsim 20~M_\odot$~\cite{Timmes:1995kp,Olejak:2019pln}. Olejak et al.~2020~\cite{Olejak:2019pln} report the most recent calculations on the population of stellar-mass black holes in the Galaxy. They find that there are $1.0 \cdot 10^8$~stellar-mass black holes in the Galactic disk. About 92\% of these black holes are isolated\footnote{Although $\sim$70\% of massive stars are in binary systems, the binary survival probability is low: compact systems may merge during the red supergiant phase of the massive star, while wide systems may be disrupted by the supernova explosion that produces the black hole.}, without any companion, while only 8\% are in binary systems. This 8\% is the sum of black holes in binaries with another black hole ($\sim$6.4\%), with a neutron star or white dwarf ($\sim$1.6\%), and with a normal star ($\sim$0.16\%). In most black hole--normal star binaries, the separation is large, making significant mass transfer from the companion star to the black hole impossible. Black holes in X-ray binaries represent only a small fraction of this 0.16\%, namely those systems that are sufficiently compact to allow the formation of a bright accretion disk from material transferred from the companion star.

Estimates of the density of stellar-mass black holes in the Galaxy are reported in Refs.~\cite{Fender:2013ei,Murchikova:2025oio,Nosirov:2026fjo}. For example, Fender et al.~2013~\cite{Fender:2013ei} model the Galactic disk as a cylinder\footnote{Modeling the Galactic disk as a cylinder of radius 10~kpc and height 0.5~kpc includes 80-90\% of the stars in the Galactic disk.} of radius 10~kpc and height 0.5~kpc. Its volume is $\sim$150~kpc$^3$. Dividing this volume by $1.0 \cdot 10^8$ -- the number of stellar-mass black holes in the Galactic disk predicted by Olejak et al.~2020~\cite{Olejak:2019pln} -- we find that there is one black hole every $\sim$1,500~pc$^3$. If we model such a volume as a sphere with a black hole at its center, the radius of this sphere is about 7~pc (approximately 23~light years). This rough estimate suggests that the closest black hole to the Solar System may lie within 7~pc, and that we may have $\sim$10~black holes within 15~pc (about 50~light years).

%%%%%%%%%%%%%%%%%%%%%%%%%%%%%%%%%%%%%%%%%%%%%%%%%%

\section{Search for Nearby Black Holes}\label{s-search} 

Since most black holes are isolated or in binary systems with another black hole or a neutron star, their detection is extremely challenging. Attempts to discover isolated black holes from microlensing events are normally possible only in very small regions of the sky characterized by a high number density of background stars, and this approach does not work for searching for a few nearby black holes ($< 15$~pc). Vladimir Shvartsman was the first to point out that isolated black holes can accrete from the interstellar medium and produce detectable electromagnetic radiation~\cite{Shvartsman}. This idea has been further explored by other authors~\cite{Meszaros75,McDowell85,Fujita:1997fh,Maccarone:2005uq,Fender:2013ei,Tsuna:2018abi,Scarcella:2020ssk,Murchikova:2025oio,Martinez:2025git}, though none had attempted to identify candidates from existing astronomical observations until the study reported in Ref.~\cite{Nosirov:2026fjo}.

\begin{figure}[t]
\centering
\includegraphics[width=0.65\linewidth]{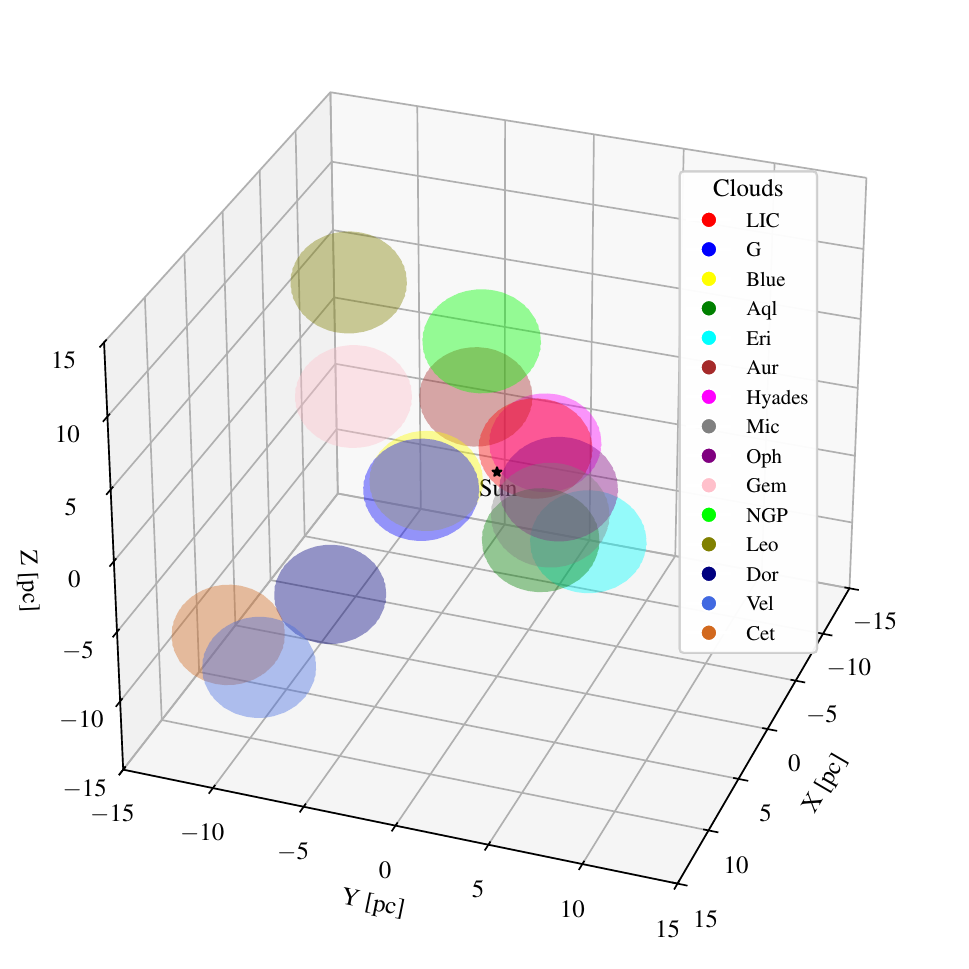}
%\vspace{-0.5cm}
\caption{Three-dimensional map of the local region within 15~pc of the Solar System. The Sun is at the center and is indicated by the black star. The Local Interstellar Clouds are modeled as spheres of radius 3~pc, and the legend reports their abbreviated names. Figure courtesy of Abdurakhmon Nosirov.}
\label{f-lics}
\end{figure}

Restricting our interest to the region within 15~pc of the Solar System, $\sim$90\% of the volume is filled with a hot, low-density interstellar medium, while a complex of warm, partially ionized clouds (the so-called Local Interstellar Clouds) occupies the remaining $\sim$10\%~\cite{Redfield:2008wm} (see Fig.~\ref{f-lics}). The temperatures of these clouds are lower than that of the hot, low-density interstellar medium, and their densities are higher. As shown in Ref.~\cite{Nosirov:2026fjo}, if a black hole were inside one of these clouds, the electromagnetic radiation produced by accretion from the interstellar medium could be detected by current and near-future observational facilities. It is possible that existing catalogs already include nearby isolated black holes accreting from the interstellar medium, but that these sources have not yet been identified as black holes. Outside these clouds, the accretion rate would be too low, and the associated accretion flux is expected to be well below the detection threshold of current and near-future catalogs.

The discovery of black holes inside these warm, partially ionized clouds, if they exist, is quite challenging. First, they would appear as faint sources with relatively featureless spectra: identification cannot rely on single-instrument observations; rather, we need observations at different wavelengths -- from the radio to the X-ray bands -- in order to rule out alternative interpretations and confirm the nature of any candidate~\cite{Murchikova:2025oio,Nosirov:2026fjo}. Second, there are large uncertainties in the exact spectra of such systems, which makes the selection of good candidates more difficult.

In Ref.~\cite{Nosirov:new}, we propose a method to find candidate isolated black holes in nearby clouds. We start from the all-sky IR catalog CatWISE2020 ($\sim$1.8~billion sources) and select faint sources with high proper motion. We then require that these sources are also present in the optical catalog NSC~DR2. At this point, for every source we have a flux measurement (or an upper bound thereof) in the W1 and W2 bands of CatWISE2020 and in the u, g, r, i, z, Y, and VR bands of NSC~DR2. These sources are mainly brown dwarfs, low-mass stars, and white dwarfs. Their spectra are well known, and there are advanced public models to fit them. According to Ref.~\cite{Nosirov:2026fjo}, the flux in the IR and optical bands of nearby black holes accreting from the interstellar medium should decrease as the frequency increases. We thus attempt to remove all brown dwarfs, low-mass stars, and white dwarfs from the available measurements in the W1, W2, u, g, r, i, z, Y, and VR bands, and we arrive at a shortlist of sources. Because of the uncertainties in the spectra of black holes accreting from the interstellar medium, we cannot make any robust prediction of their fluxes in the radio and X-ray bands, which may lie above or below the detection threshold of current radio and X-ray catalogs such as VLASS~QL1, RACS-mid, RACS-high, or eROSITA. A multi-wavelength campaign targeting the most promising candidates appears to be the only method to further test the nature of these sources and possibly discover some nearby isolated black holes.

%%%%%%%%%%%%%%%%%%%%%%%%%%%%%%%%%%%%%%%%%%%%%%%%%%

\section{Interstellar Mission}\label{s-m}

The closest star to the Solar System is Proxima Centauri, which lies 4.24~light-years from the Sun. It is thus clear that exploration beyond the Solar System requires the development of spacecraft that can travel at some significant fraction of the speed of light. Current chemically propelled spacecraft cannot achieve this, as can be readily seen from the Tsiolkovsky rocket equation, $\Delta v = v_e \ln \left( m_i / m_f \right)$, where $\Delta v$ is the total change of the rocket's velocity, $v_e$ is the effective exhaust velocity of the propellant gases relative to the rocket, and $m_i$ and $m_f$ are the initial total mass (with propellant) and the final total mass (without propellant) of the rocket, respectively. Today, the most efficient chemical propellant is a combination of liquid hydrogen and liquid oxygen, with $v_e \sim 4.5$~km/s: even if $m_f$ were the proton mass and $m_i$ were the mass of the Earth, $\Delta v = 0.0018~c$, where $c$ is the speed of light, and our rocket would take over 2,000~years to reach Proxima Centauri.

The study of space missions outside the Solar System can be traced back to the Orion project in the 1950s~\cite{Orion} and the Daedalus project in the 1970s~\cite{Daedalus}. For the ``near'' future (i.e., within the next few decades), the most promising solution for interstellar exploration is currently thought to be laser propulsion, which was originally proposed in the 1960s~\cite{Marx,Redding}. Over the last 10-20 years, the exoplanet community has repeatedly discussed the possibility of sending ultralight probes to explore exoplanets orbiting nearby stars~\cite{Lubin16,Lubin22,Parkin18,Kuhlmey25,Eubanks2026}. The most famous program is likely the Breakthrough Starshot Initiative\footnote{\href{https://breakthroughinitiatives.org/}{https://breakthroughinitiatives.org/}}, which had the goal to reach Alpha Centauri in about 20~years with a spacecraft capable of traveling at one-fifth the speed of light, but was discontinued in September 2025.

\begin{figure}[t]
\centering
\includegraphics[width=0.95\linewidth]{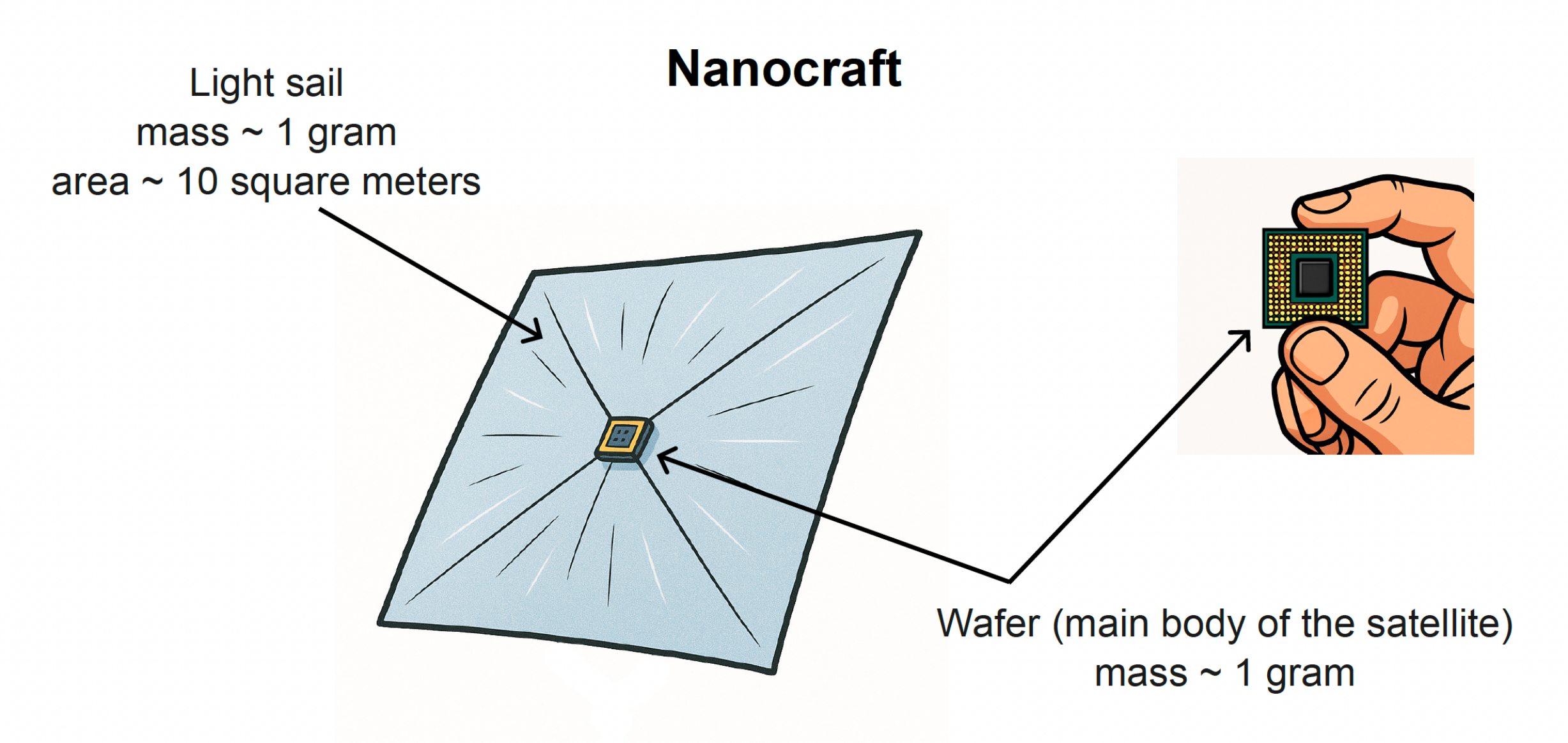}
\vspace{-0.3cm}
\caption{Sketch of a nanocraft. Figure from Ref.~\cite{Bambi:2025kcr}. 
\label{f-n}}
\end{figure}

In the framework of laser propulsion, the concept of spacecraft is quite different from the traditional idea. The spacecraft is often called a {\it nanocraft} because of its very low mass; a sketch is shown in Fig.~\ref{f-n}. There are two main parts: a gram-sized wafer, containing a computer processor, navigation, and communication systems, and a meter-scale, extremely thin, dielectric metamaterial light sail. High-power lasers strike the light sail, and the resulting radiation pressure accelerates the nanocraft to its target speed. There are no fundamental barriers preventing the attainment of 90\% of the speed of light using this approach, although higher velocities immediately increase mission costs. Since the nanocraft does not carry any fuel (the lasers are located on the ground or in space), it is not subject to the Tsiolkovsky rocket equation.

\begin{figure}[t]
\centering
\includegraphics[width=0.95\linewidth]{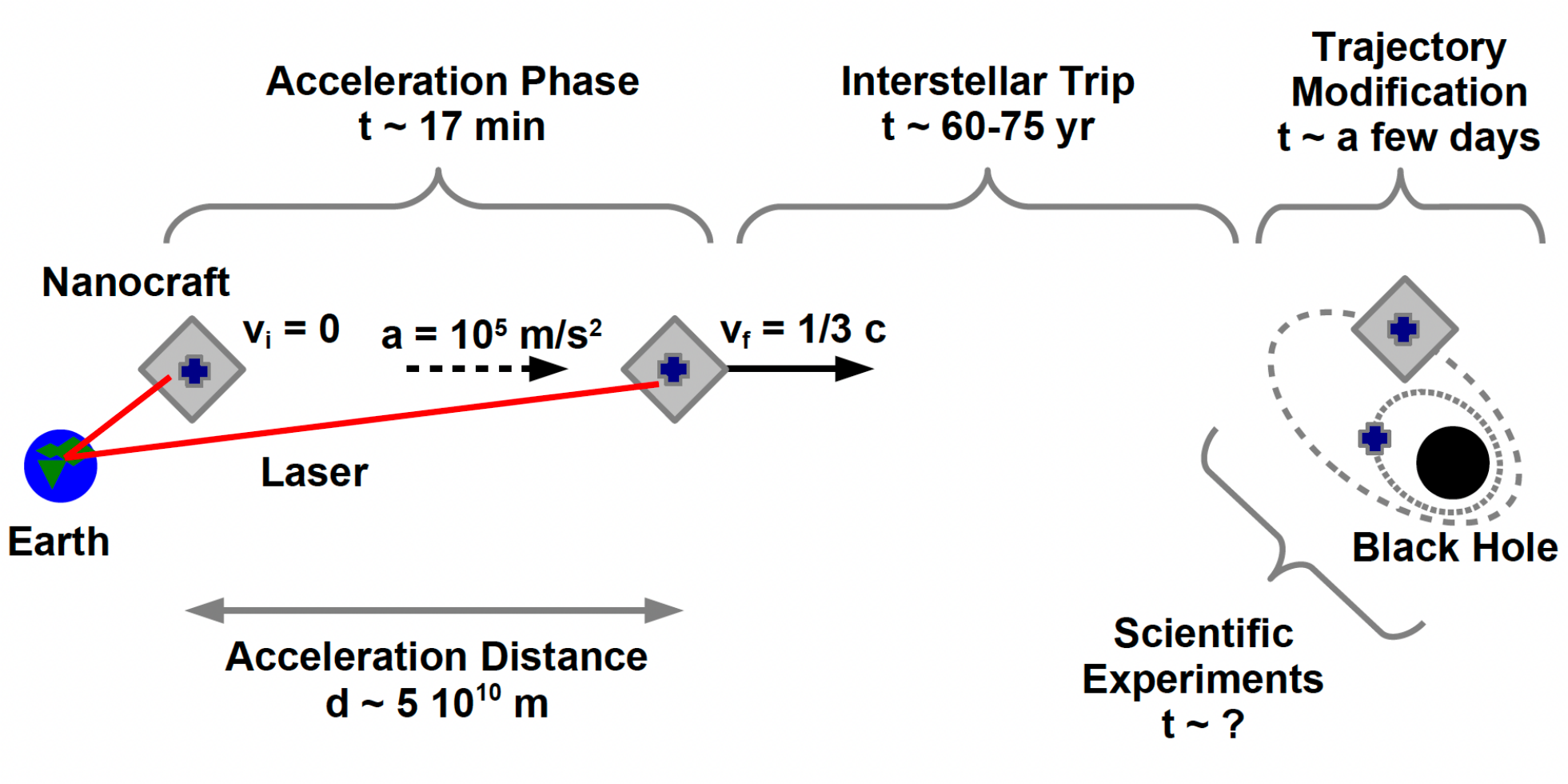}
\vspace{-0.3cm}
\caption{Phases of a hypothetical interstellar mission with a nanocraft to the closest black hole. Figure from Ref.~\cite{Bambi:2025kcr}. 
\label{f-m}}
\end{figure}

If we can find a black hole at 20-25~light years from the Solar System and can develop nanocrafts capable of traveling at one-third the speed of light, we can consider the possibility of an interstellar mission lasting 80-100~years: the nanocrafts need 60-75~years to reach the black hole, they test the nature of the compact object, and they send the data back to Earth. The latter take 20-25~years to reach the Solar System, so the overall mission would span almost a century~\cite{Bambi:2025kcr,Bambi:2025hjn}.

Fig.~\ref{f-m} illustrates the phases of such an interstellar mission to the closest black hole. Assuming that the maximum acceleration of a nanocraft is $a_{\rm max} \approx 10^5$~m~s$^{-2}$ and the target velocity is one-third the speed of light, we need $10^3$~s ($\sim$17~minutes) for the acceleration phase at maximum acceleration. After that, the lasers are turns off, and the nanocraft can begin its interstellar journey: if the black hole is at 20-25~light years from the Solar System, the trip takes 60-75~years. When the mission reaches the black hole, the probe(s) should conduct all scientific experiments (see the next section) and eventually send the data back to Earth. The data will arrive after 20-25~years, so the total duration of the mission may be 80-100~years.

%%%%%%%%%%%%%%%%%%%%%%%%%%%%%%%%%%%%%%%%%%%%%%%%%%

\section{Experiments Near the Black Hole}\label{s-exp}

In 4D~General Relativity and in the absence of exotic matter fields, the final product of gravitational collapse should be a Kerr black hole, which is characterized solely by its mass and spin angular momentum\footnote{Here we ignore a possible non-vanishing electric charge because the equilibrium electric charge should be negligible for a stellar-mass compact object, and we do not know of any astrophysically viable mechanism to create a stellar-mass black hole with a significant electric charge; see, for instance, Refs.~\cite{Bambi:2017khi,Bambi:2008hp}.}. An interstellar mission to a black hole can thus address the following questions. Is the spacetime geometry around an astrophysical black hole well described by the Kerr solution\footnote{The mass of the accreting material around the black hole is many orders of magnitude lower than the mass of the black hole, and its impact on the spacetime geometry is completely negligible even for the desired very precise and accurate measurements of our interstellar mission; see, for instance, Ref.~\cite{Bambi:2014koa} for some estimates of the impact of the accretion material on the spacetime metric.}? Do astrophysical black holes have an event horizon?

To test the Kerr nature of the black hole, we need to study the motion of massive or massless particles around the compact object. We need at least two probes that can communicate with each other. Preliminary studies on how we can test the spacetime geometry around a black hole are reported in Refs.~\cite{Gao:2026jpl,Fan:new}. At the moment, there is no clear solution for how to decelerate a nanocraft after it reaches its target source (black hole or other). The study in Ref.~\cite{Gao:2026jpl} assumes that we have the ability to decelerate the nanocrafts and shows that this alone is not sufficient: we need at least one probe to orbit near the innermost stable circular orbits in order to perform very precise tests of the Kerr metric. The study in Ref.~\cite{Fan:new} considers the opposite situation: we cannot decelerate the nanocrafts and must instead perform flyby experiments. It turns out that very precise and accurate flyby experiments are possible if we have the capability of sending some probes near the critical radii that separate trajectories returning to infinity from those that fall onto the black hole. The conclusion of Ref.~\cite{Fan:new} is that tiny probes with a size of $\sim$1~cm may potentially test deviations from the predictions of General Relativity beyond the first post-Newtonian order with an accuracy at the level of $10^{-6}$. This number should be compared with current constraints from X-ray and gravitational-wave data at the level of $\sim$0.1 (see, for instance, Ref.~\cite{Das:2026zyt}) and future constraints from space-based laser interferometers (such as LISA, which is currently expected to operate around 2037) at the level of $\sim$0.01.

%%%%%%%%%%%%%%%%%%%%%%%%%%%%%%%%%%%%%%%%%%%%%%%%%%

\section{Concluding Remarks}\label{s-cr}

The idea of an interstellar mission to Earth's nearest black hole is certainly extremely ambitious: we do not have the technology today, and there are many problems to understand and solve. However, none of these problems appears to be impossible to solve at the moment. For example, so far there are no clear ideas of how nanocrafts can decelerate once they reach their target source, and this problem has been frequently discussed even by the exoplanet community in the context of exploring exoplanets in nearby stellar systems. However, as discussed above, we may not need to decelerate the nanocrafts to test the nature of a black hole: flyby experiments can achieve the same goal if we can get sufficiently close to the black hole.

While it will certainly take some time to reach the technology necessary to send a probe to a nearby black hole, there are many intermediate steps that appear extremely valuable in their own right. For example, one of the first steps is to find a black hole within 20-25~light-years of the Solar System. This would already be a breakthrough in physics and astrophysics, because we do not currently know of any black hole accreting from the interstellar medium; such a discovery would therefore represent a completely new kind of astrophysical source.

The development of nanocrafts for exploration beyond the Solar System could be extremely useful for many research areas, not only for black holes and tests of General Relativity. We can indeed imagine that the first probes could explore the boundaries of the Solar System and the interstellar medium, without requiring the capability of reaching a very specific source or possessing very advanced navigation systems. A second generation of probes may explore objects just outside the Solar System, such as nearby exoplanets and stellar systems. Only when this technology is sufficiently mature could we undertake an interstellar mission to a black hole, assuming that we find one sufficiently close to us.

Apart from the search for nearby black holes, which is our current priority, our efforts are dedicated to creating a global consortium for the development of these nanocrafts for exploration beyond the Solar System. We are establishing four working groups: science (WG1), probe and instrumentation (WG2), launch (light sail and laser, WG3), and communication (WG4). We hope to have some preliminary working groups active soon and to organize an international meeting to discuss the corresponding science and technology in Summer 2027.

%%%%%%%%%%%%%%%%%%%%%%%%%%%%%%%%%%%%%%%%%%%%%%%%%%

\acknowledgments

This work was supported by the National Natural Science Foundation of China (NSFC), Grant No.~W2531002.

%%%%%%%%%%%%%%%%%%%%%%%%%%%%%%%%%%%%%%%%%%%%%%%%%%

\bigskip
\bigskip
\noindent {\bf DISCUSSION}

\bigskip
\noindent {\bf ANDREA GOKUS:} How much do the orbital motions of a black hole with respect to the Solar System need to be taken into account? Will it be possible to achieve the accuracy needed for a nanocraft to get sufficiently close to a stellar-mass black hole?

\bigskip
\noindent {\bf COSIMO BAMBI:} Precise and accurate determinations of the position and velocity of the target black hole are certainly very important for the success of the mission, especially in the case of a flyby experiment. The details depend on the design of the mission. In any case, we can expect that it will be unlikely to measure these quantities with the necessary precision using observations from Earth alone. For example, we might have a first swarm of probes that reach the black hole earlier, measure its position and velocity, and communicate these measurements to a second swarm of probes that arrive later and test the compact object. We could also consider complementary solutions, such as having the nanocrafts (with masses of a few grams and sizes of a few meters) launched from Earth release -- when they are near the target black hole -- a large number of tiny (sub-gram and sub-cm) probes that can get very close to the black hole. It is not necessary that all these small probes get sufficiently close; we only need a few of them to do so, and those probes can then perform the experiments required to test the nature of the compact object.

\bigskip
\bigskip
\noindent {\bf ALEXANDER DOLGOV:} Are you considering only black holes formed from the gravitational collapse of stars? Could these black holes be primordial?

\bigskip
\noindent {\bf COSIMO BAMBI:} The claim that there may be one black hole within 7~pc of the Solar System, and $\sim$10~black holes within 15~pc, assumes that there are only black holes formed from the gravitational collapse of massive stars. This estimate should be relatively reliable because it is based on standard physics. If primordial black holes exist, their mass distribution and local density are very model-dependent, so it is difficult to make predictions. For example, if their masses were lower than $\sim$10~$M_\odot$, the radiation produced by accretion from the interstellar medium would be weaker than our predictions for stellar-mass black holes, and their detection and identification would be difficult unless these objects are much closer to us. However, if we were to discover a nearby primordial black hole, I do not see any substantial difference from the mission perspective, regardless of whether the black hole is primordial or not.

\end{document}